\documentclass{amsart}
\usepackage{graphicx} % Required for inserting images
\usepackage{amsthm}
\usepackage{ragged2e}
\usepackage{blindtext}
\usepackage{enumerate}
\usepackage{amssymb}
\usepackage{amsfonts}
\usepackage{latexsym}
\usepackage{amsmath}
\usepackage{euscript}
\usepackage{graphics}
\usepackage{graphicx}
\usepackage{tikz}
\usetikzlibrary{spy}
\graphicspath{ {./images/} }

\newtheorem{theorem}{Theorem}[section]

\newtheorem{claim}[theorem]{Claim}

\title[Early State Exclusion for $5\times5$ Jacobi Matrices]{Early State Exclusion for $5\times5$ Jacobi Matrices }
\author{Jasmin Enriquez, Anastasiia Minenkova }
\address{
Department of Mathematics\\
University of Hartford\\
200 Bloomfield Avenue\\
West Hartford, CT 06117, USA}
\email{jenriquez@hartford.edu, minenkova@hartford.edu}
\date{\today}

\begin{document}

\maketitle
\begin{abstract}
In this paper, we present a problem that arose from quantum computing, when the quantum state to be transferred completely vanishes before the first occurrence of the perfect state transfer. It was posed as an open problem for 1D quantum chains with $n$ of qubits, where $n>4$ is an odd number. We present the solution for $n=5$. The proofs are accessible to students who studied trigonometry and calculus. As for the set-up the knowledge of eigentheory and inner products from linear algebra is needed.

\end{abstract}



\section{Introduction}

Transmitting a quantum state is important for connecting small quantum processors to create large-scale quantum computers. To do this, we need physical systems that can act as channels for quantum communication. We can either send the state directly or use the channel to share entanglement and then teleport the state.
For long-distance communication, optical fibers are ideal because they can connect quantum computers with optical systems. For short-distance communication, like between nearby quantum processors, other methods are being explored, such as using ion traps. The study of state transfer was initiated by Bose (see \cite{B03})
who considered a 1D chain of $N$ qubits (open boundary conditions) coupled by the time-independent Hamiltonian.

Since then, numerous researchers (see \cite{Kay10,VZh12,Kall} and many more) have made significant contributions. Their work has boosted various areas of mathematics, such as graph theory and numerical analysis. Problems like perfect state transfer and pretty good state transfer have garnered interest from both the mathematics community and the field of quantum information.

In the present paper, we consider the problem of early state exclusion. In \cite{ESE}, the authors showed the construction of 1D Hamiltonians that realize perfect quantum state transfer (PST) and also have the property that the overlap of the time-evolved state with the initial state is zero for some time before the first PST time. When the latter takes place, they called it an early state exclusion (ESE). The authors showed infinite families of Hamiltonians realizing PST between end-vertices of the weighted path with (in the case of an even number of qubits in the chain) and without ESE (in the case of three qubits in the chain or in the case of the equidistant spectrum). We aim to extend this study to the case of 5 qubits and to show the construction in which early state exclusion occurs.

The 1D Hamiltonian of the system of $N+1$ qubits with nearest-neighbor interactions is represented by a Jacobi matrix $J$, which is a symmetric tridiagonal matrix of the form
\[
J=\begin{bmatrix}%{cccc}
		a_0 & b_0    &  &  \\
		b_0 & a_1    & \ddots &  \\
		& \ddots & \ddots & b_{N-1} \\
		&        & b_{N-1}& a_{N}     \\
	\end{bmatrix},
	\]
where the $a_k$ are real numbers and the $b_k$ are positive. The time evolution of the system is then given by $e^{-itJ}$.

We consider Jacobi matrices with centered spectrum, that is, if $\lambda_0<\lambda_1<\lambda_2<\ldots<\lambda_N$ are the eigenvalues of $J$, then
\[
\lambda_k-\lambda_{k-1}=\lambda_{N-k}-\lambda_{N-k+1}, \text{ for }k=1,\ldots, N-1.
\]

\subsection{Perfect Quantum State Transfer}

A matrix $H$ realizes \emph{perfect state transfer} (PST) between the end-vertices of the weighted path if there exist some positive $T_0$ and real $\phi$ for which
\begin{equation}\label{PST}
e^{-iT_0H}{\bf e}_0=e^{i\phi}{\bf e}_N,
\end{equation}
where ${\bf e}_0$ and ${\bf e}_N$ are the first and last elements of the standard basis
$\{ {\bf e}_k\}_{k=0}^N$ of $\mathbb{C}^{N+1}$.

To determine if PST between end-vertices of the weighted path is possible in a 1D nearest-neighbor coupled chain, one must check a symmetry condition and then examine the eigenvalues of an $(N \times N)$ matrix.
That is, PST between end-vertices of the weighted path is realized when a Jacobi matrix $J$ has the following two properties (see \cite{DMS,Kay10,VZh12}):
\begin{enumerate}
\item $J$ is persymmetric, that is,
\[
\begin{split}
a_{k}=a_{N-k},\quad k=0,1,\dots, N, \\
 b_{l}=b_{N-1-l}, \quad l=0,1,\dots, N-1;    
\end{split}
\]
\item the eigenvalues $\lambda_0<\lambda_1<\dots<\lambda_N$ of $J$ satisfy
\begin{equation}\label{PST_eigenvalues}
    \lambda_{k+1}-\lambda_k=\frac{(2n_k+1)\pi}{T_0}, \quad k=0,...,N-1,
\end{equation}
where $n_k$ are nonnegative integers.
\end{enumerate}
%As for the transfer time $T_0$, it is subject to the following constraint: The minimum spacing between the eigenvalues is a rational multiple of $\frac{\pi}{T_0}$.

\subsection{Early State Exclusion}

In \cite{ESE}, the authors introduced the following definition of early state exclusion.
Let $J$ be a Jacobi matrix that has the earliest PST between end-vertices of the weighted path occurring at time $T_0$. If there is a time $t$ such that $0<t<T_0$ and
\[
\langle e^{-iJt}{\bf e}_0,{\bf e}_0\rangle_{\mathbb{C}^{N+1}} = 0
\]
then $J$ has an $\textit{Early State Exclusion}$ (ESE) at time $t$. The ESE phenomenon is not necessarily unique and could happen a few times (see Figure 1).
\begin{figure}[h]
\centering
\includegraphics[width=0.6\textwidth]{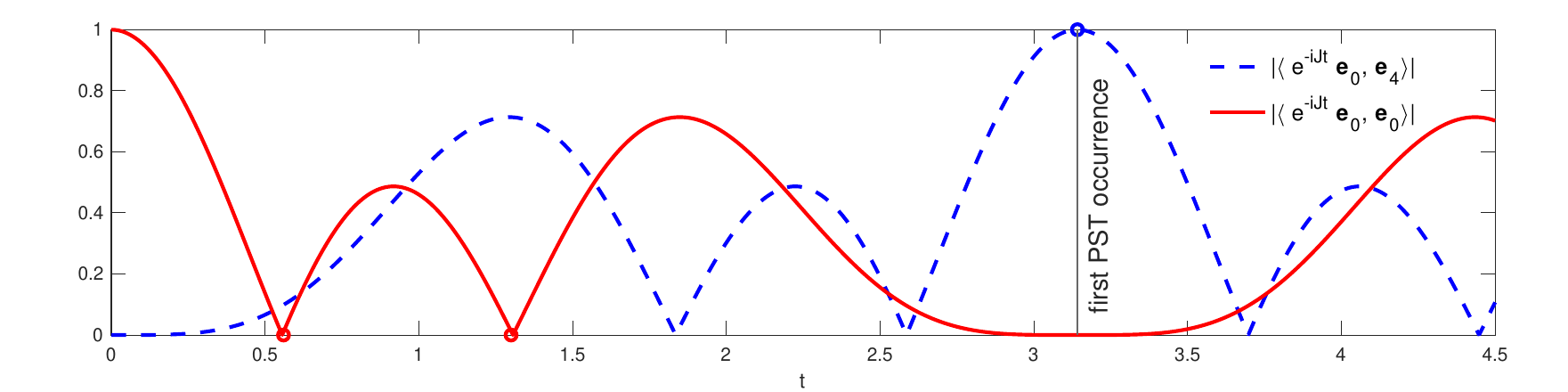}
\caption{Example of ESE occurrences (indicated by the two dots on the solid line representing $|\langle e^{-iJt}{\bf e}_0,{\bf e}_0\rangle_{\mathbb{C}^{N+1}}|$) before the first PST occurrence.}
\label{FigESE34}
\end{figure}

\subsection{Main Result}
After numerical exploration, the results of which the reader will see throughout the paper, we observed that ESE occurred in some cases of $5\times5$ Jacobi matrices with a centered spectrum.
This ultimately culminated in the discovery of the following result. We present the analytical proof of this statement in Section~3.

\begin{theorem}\label{main}
There are infinitely many $5\times5$ symmetric and persymmetric Jacobi matrices realizing PST between end-vertices of the weighted path that have ESE and infinitely many that do not. Namely, the $5\times5$ Jacobi matrices with centered spectrum $\{c-x,c-y,c,c+y,c+x\}$ ($x>y>0$ and $c\in\mathbb{R}$) realizing PST between end-vertices of the weighted path:
 do not have ESE if and only if $x$ is an integer multiple of $y$.
\end{theorem}

\section{ESE for Persymmetric Jacobi Matrices with Symmetric Spectra}

Before considering a case with centered spectrum, let us consider a slightly simplified version.

Let $\lambda_0<\lambda_1<\lambda_2<\ldots<\lambda_N$ be the eigenvalues of $J$.
We say that $\{\lambda_k\}_{k=0,\ldots,N}$ is the symmetric spectrum if 
\[
\lambda_k=-\lambda_{N-k}, \text{ for }k=0,\ldots, N.
\]

Let us consider the case of $J$ being a $5\times5$ Jacobi matrix the symmetric spectrum $\{-x,-y,0,y,x\}$ for some positive $y<x$.
If $J$ realizes PST between end-vertices of the weighted path, it is persymmetric and is uniquely defined by its spectrum due to Theorem 3 in \cite{H74}. So it is of the form
\begin{equation}\label{J5}
J = \begin{bmatrix}
0 & y&0&0 & 0 \\
y&0 & \sqrt{\frac{x^2-y^2}{2}} & 0&0 \\
0 & \sqrt{\frac{x^2-y^2}{2}} &0 & \sqrt{\frac{x^2-y^2}{2}} & 0& \\
0&0 & \sqrt{\frac{x^2-y^2}{2}} & 0&y \\
0 & 0 & 0&y&0
\end{bmatrix}
\end{equation} 
 and from \eqref{PST_eigenvalues} we have \begin{equation}\label{eq:xny}
     y=\frac{\pi(2n+1)}{T_0}\text{ and }x=\frac{2m\pi
 }{T_0}
 \end{equation} for some nonnegative integers $n,m$, where $T_0$ is the first time the PST between end-vertices of the weighted path occurs. The algorithm for the persymmetric matrix reconstruction from its spectrum can be found in \cite{BG,HM}.

\begin{theorem}
    Let $J$ be a $5\times 5$ Jacobi matrix with a symmetric spectrum such that PST between end-vertices of the weighted path occurs. 
    Then $J$ does not have ESE if one of its positive eigenvalues is an integer multiple of another positive eigenvalue and $J$ has ESE otherwise.
\end{theorem}
    
We will divide this theorem into two  parts: the case when a positive eigenvalue (Claim~\ref{ClaimNoESE}) is an integer multiple of another positive eigenvalue of $J$ and the case when it is not (Claim~\ref{ClaimESE}).

From now on assume that $J$ is of the form \eqref{J5} and $T_0$ is the first time PST between end-vertices of the weighted path occurs for $J$. We compute $e^{-iJt}$ using Sylvester's formula,
\begin{align*}
   e^{-iJt}=&
   \frac{e^{ixt}}{(-x+y)(
   -x)(-x-y)(-x-x)}(J+yI)J(J-yI)(J-xI)+\\
&+\frac{e^{iyt}}{(-y+x)(
   -y)(-y-y)(-y-x)}(J+xI)J(J-yI)(J-xI)+\\
&+\frac{1}{y
   x(-y)(-x)}(J+xI)(J+yI)(J-yI)(J-xI)+
\\
&+\frac{e^{-iyt}}{(y+x)(
   y+y)y(y-x)}(J+xI)(J+yI)J(J-xI)+
 \\&+\frac{e^{-ixt}}{(x+x)(
   x+y)x(x-y)}(J+xI)(J+yI)J(J-yI).
\end{align*}
For our purpose we do not actually need the entire matrix. We just need to know that
\begin{align*}
    {\bf e}_0^\top
(J+yI)J(J-yI)(J-xI){\bf e}_0&=\frac{y^2(x^2- y^2)}{2}, \\
{\bf e}_0^\top(J+xI)J(J-yI)(J-xI){\bf e}_0&=\frac{y^2(y^2- x^2)}{2},\\
{\bf e}_0^\top
(J+xI)(J+yI)(J-yI)(J-xI){\bf e}_0&=\frac{y^2(x^2- y^2)}{2},
\\
{\bf e}_0^\top(J+xI)(J+yI)J(J-xI){\bf e}_0&=\frac{y^2(y^2- x^2)}{2},\\
{\bf e}_0^\top(J+xI)(J+yI)J(J-yI){\bf e}_0&=\frac{y^2(x^2- y^2)}{2}.
\end{align*}
Then \begin{equation*}
    \langle e^{-iJt}{\bf e}_0,{\bf e}_0\rangle_{\mathbb{C}^{N+1}} =\frac{1}{2x^2}( x^2-y^2+x^2\cos(yt)+y^2\cos(xt))\overset{\text{\eqref{eq:xny}}}{=}
       \end{equation*}
\begin{equation}\label{expJ}
=\frac{1}{8m^2}\left((2m)^2-(2n+1)^2+(2m)^2\cos\left(\frac{(2n+1)t\pi}{T_0}\right)
    +(2n+1)^2\cos\left(\frac{2mt\pi}{T_0}\right)\right).
\end{equation}
Note that this is a real-valued function.

\subsection{Case when $x$ is a multiple of $y$}

While running simulation on MATLAB, we have noticed that certain cases did not have ESE (see Figure 2). First, we have noticed that the Jacobi matrices with the equidistant eigenvalues do not have ESE. The same picture appeared when $x$ was an integer multiple of $y$.

\begin{figure}[h]\label{NoESE}
\caption{Examples of $J$ without ESE with eigenvalues $\{-2,-1,0,1,2\}$ in the first graph, $\{-6,-1,0,1,6 \}$ in the second graph, and $\{-12,-2,0,2,12\}$.}
\centering
\includegraphics[height=0.125\textheight]{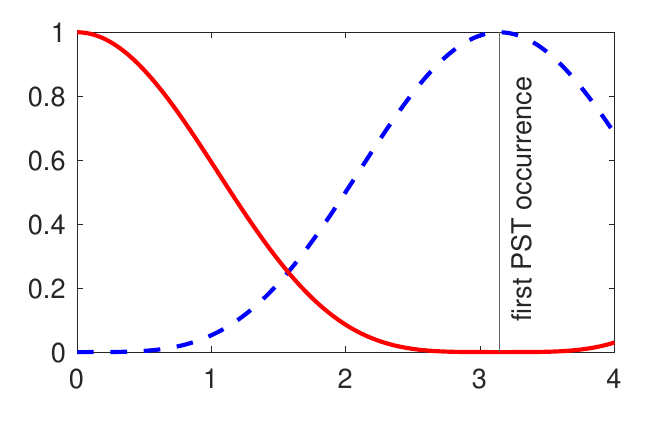}\includegraphics[height=0.125\textheight]{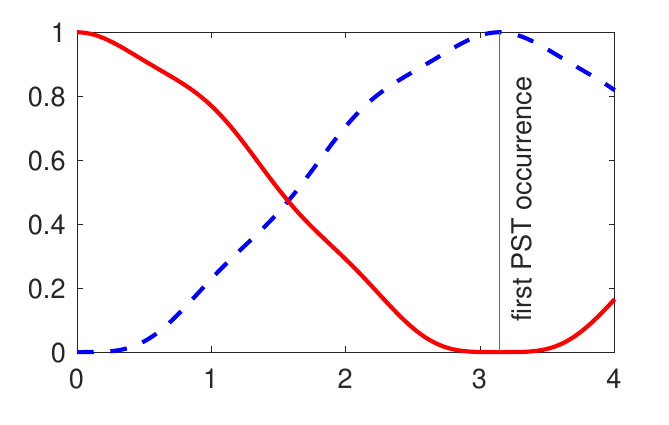}
\includegraphics[height=0.125\textheight]{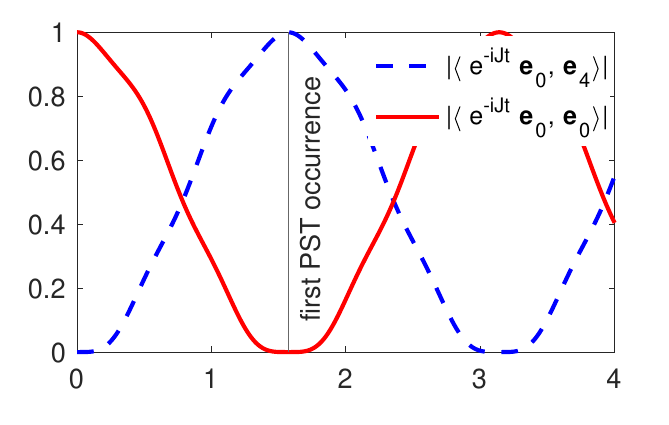}
\end{figure}

This observation led us to discover the following fact.
\begin{claim}\label{ClaimNoESE}
    A $5\times 5$ Jacobi matrix with a symmetric spectrum realizing PST between end-vertices of the weighted path does not have ESE if one of its positive eigenvalues is an integer multiple of another positive eigenvalue.
\end{claim}

\begin{proof}
    Without loss of generality, we can assume that the given matrix is $J$ from \eqref{J5}, $T_0=\pi$, $y=1$, and $x=2m$ for some positive integer $m$ in \eqref{expJ}, that is
\begin{equation}\label{expJmultiple}
    \langle e^{-iJt}{\bf e}_0,{\bf e}_0\rangle_{\mathbb{C}^{N+1}} = (2m)^2-1+(2m)^2\cos\left(t\right)+\cos({2mt}).
\end{equation}
The ESE condition means that this function must have a zero at $t<\pi$, i.e.
\begin{equation*}
     (2m)^2-1+(2m)^2\cos\left(t\right)+\cos({2mt})=0
\end{equation*}
or
\begin{equation*}
(2m)^2(1+\cos\left(t\right))=1-\cos({2mt}).
\end{equation*}
Let us study both sides of the last equation separately. Let $$f(t)=(2m)^2(1+\cos\left(t\right))$$ and $$g(t)=1-\cos({2mt}).$$
We claim that $f(t)\ge g(t)$ on $[0,\pi]$ (see Figure~\ref{fig:FvsG}) with equality only at $t=\pi$, which would imply that \eqref{expJmultiple} does not have zeros on $[0,\pi)$. 
\begin{figure}[h]
  \begin{center}
\includegraphics[height=0.15\textheight]{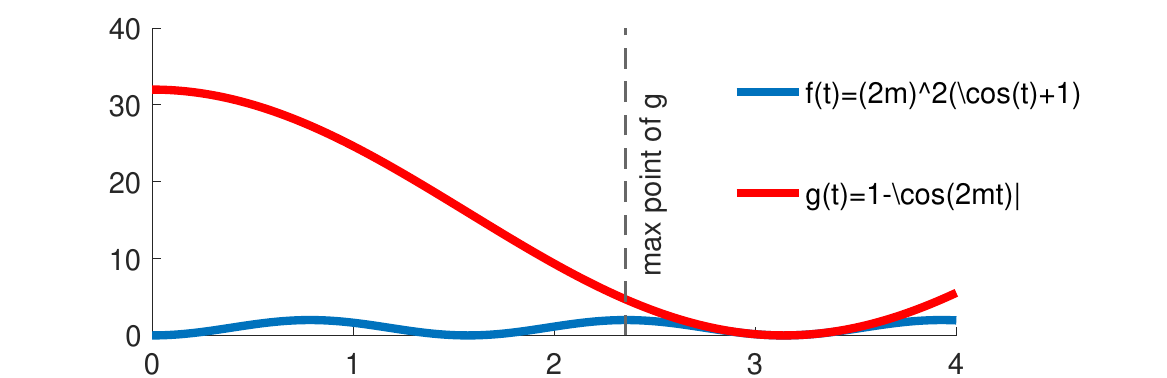}
\end{center}
    \caption{An example of  $f$ and $g$ when $m=2$.}
    \label{fig:FvsG}
\end{figure}

We start with comparing these functions and their derivatives.
\begin{center}
\begin{tabular}{c ||c |c} 
 \hline
  & $f$& $g$  \\ [0.5ex] 
 \hline\hline
 original function&$(2m)^2(1+\cos\left(t\right))$&$1-\cos({2mt})$\\ [1ex] \hline
 $\frac{d}{dt}$& $-(2m)^2\sin\left(t\right)$ & $2m\sin({2mt})$ \\ [1ex] 
 \hline
$\frac{d^2}{dt^2}$ & $-(2m)^2\cos\left(t\right)$ & $(2m)^2\cos({2mt})$ \\ [1ex] 
 \hline
\end{tabular}
\end{center}

%Instead of considering $[0,\pi]$ we consider $[\pi,2\pi]$ and the rest will follow by symmetry and periodicity of cosine. 

Note that $f(\pi)=g(\pi)=0$, $\frac{df}{dt}(\pi)=\frac{dg}{dt}(\pi)=0$, $\frac{df}{dt}>0$ on $[0,\pi)$, and $g$ attains its last maximum on $[0,\pi]$ at $t=\pi-\frac{\pi}{2m}$ and never exceeds it on this segment while values of $f$ are getting greater (and at some point greater than 1) as $t$ decreases from $\pi-\frac{\pi}{2m}$ to 0. Hence, if we show  $\frac{d(f-g)}{dt}<0$ on $[\pi-\frac{\pi}{2m},\pi]$ this will conclude the proof.

Notice that after the last maximal point before $t=\pi$ for $\frac{d^2g} {dt^2}$,  $\frac{d^2f}{dt^2}$ increases slower than $\frac{d^2g} {dt^2}$ due to the frequency in the arguments, and $\frac{d^2(f-g)}{dt^2}(\pi)=0$. Thus, $\frac{d^2(f-g)}{dt^2}>0$ from the previous zero of $\frac{d^2(f-g)}{dt^2}$ to $t=\pi$. To find this point we need to solve the following equation.
\[
\frac{d^2(f-g)}{dt^2}=-(2m)^2\cos\left(t\right)-(2m)^2\cos({2mt})=0\text{ or } \cos(t)=-\cos(2mt).
\]
{That is, }$
t=\pi\pm2mt+2\pi k,\text{ where }k\in\mathbb{Z.}$

So the point we are looking for is $t=-\frac{2\pi}{2m+1}+\pi<\pi-\frac{\pi}{2m}$. This means that $\frac{df}{dt}$ increases faster than $\frac{dg}{dt}$ on $[\pi-\frac{\pi}{2m},\pi]$, and since $\frac{df}{dt}(\pi)=\frac{dg}{dt}(\pi)$ it means that $\frac{d(f-g)}{dt}<0$ on $[\pi-\frac{\pi}{2m},\pi]$, and hence on $[0,\pi]$. Therefore, $f-g$ does not vanish on $(0,\pi)$ and $J$ does not have ESE. The general result follows the same lines after rescaling.
\end{proof}

\subsection{Case when $x$ is not a multiple of $y$}
When $x$ is not an integer multiple of $y$ we saw the following behavior.
\begin{figure}[h]\label{FigESEs}
\caption{Examples of $J$ realizing PST with ESE.}
\begin{flushright}
\includegraphics[height=0.111\textheight]{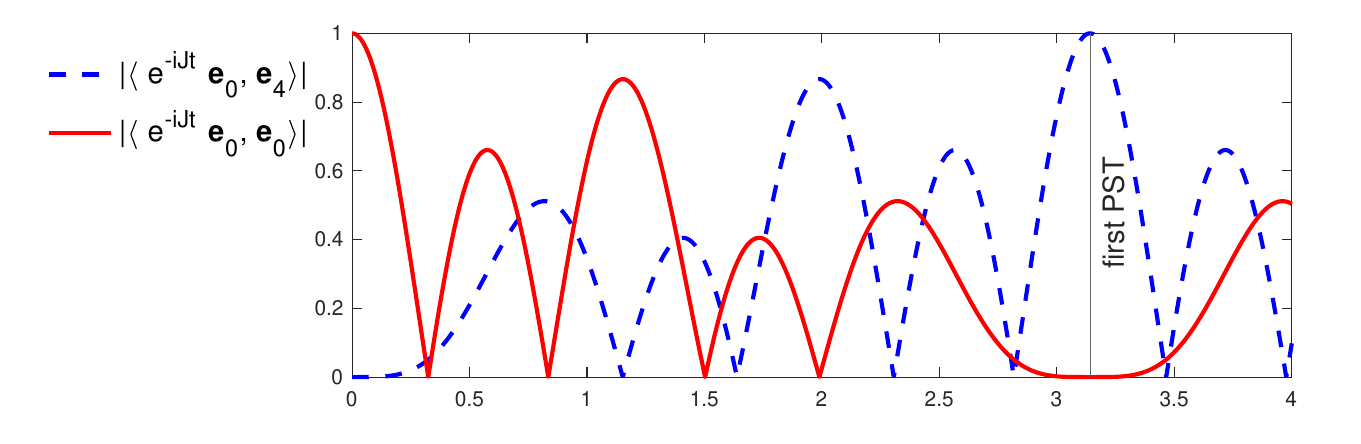}\includegraphics[height=0.111\textheight]{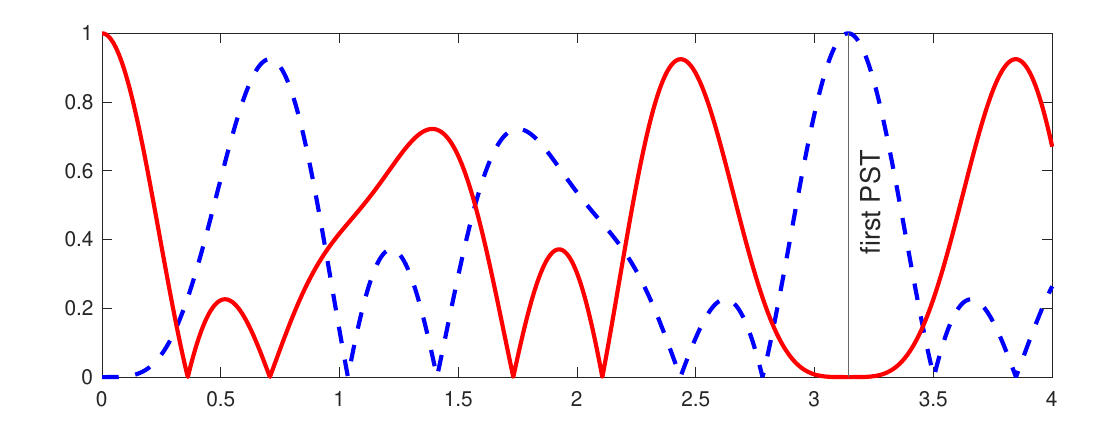}\hfill

\hfill\hfill$x=6,\ y=5$\hfill\hfill$x=8,\ y=5$\hfill\hfill
\vskip 5pt

\includegraphics[height=0.112\textheight]{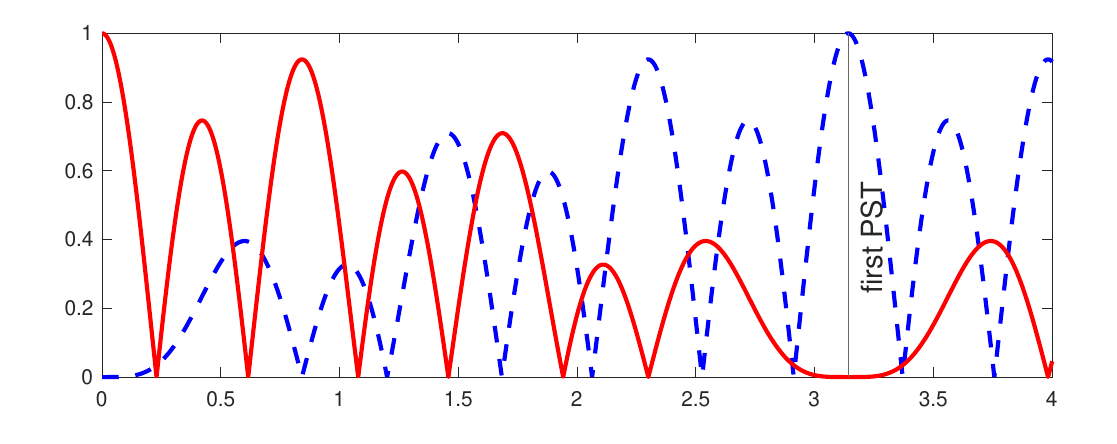}\includegraphics[height=0.112\textheight]{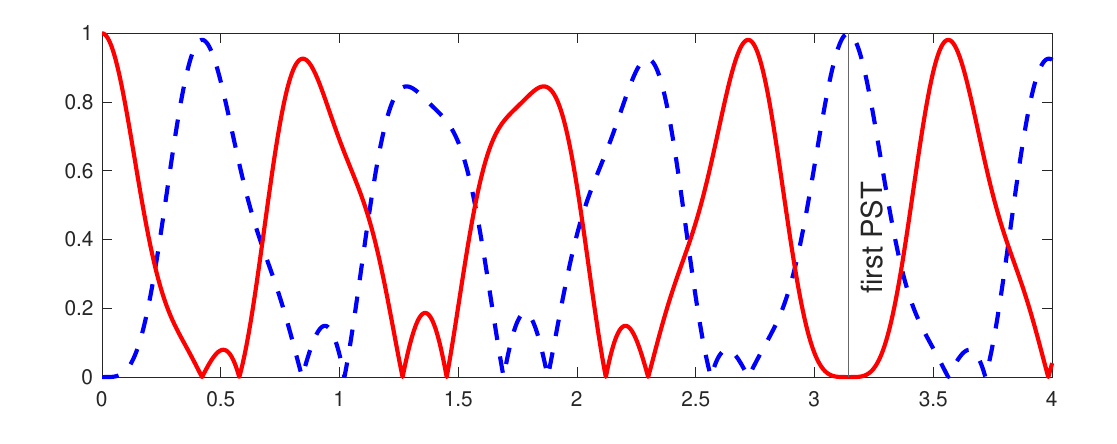}

\hfill\hfill$x=8,\ y=7$\hfill\hfill$x=16,\ y=7$\hfill\hfill

\end{flushright}
\end{figure}

Moreover, we were able to find a Jacobi matrix having any given even number $2k$ of the ESE occurrences before the transfer time $T_0$ by taking $y=2k+1$ and $x=2k+2$.

\begin{claim}\label{ClaimESE}
    A $5\times 5$ Jacobi matrix with a symmetric spectrum realizing PST between end-vertices of the weighted path has ESE if its positive eigenvalues are not integer multiple of each other. Moreover, there is a Jacobi matrix with the prescribed even number of ESE occurrences.
\end{claim}

\begin{proof}
    Without loss of generality, assume that $T_0=\pi$ and $x=2m$ and $y=2n+1$ are coprime ($x>y$), where $n$ and $m$ are some positive integers. If the numbers are not coprime, we can rescale $T_0$ and consider an equivalent rescaled problem. 
    
    By combining \eqref{expJ}  with the definition of ESE we get
    \begin{equation}\label{EqESE}
(2m)^2-(2n+1)^2+(2m)^2\cos\left({(2n+1)t}\right)
    +(2n+1)^2\cos\left(2mt\right)=0.
\end{equation}
As before, we consider two functions instead: $f(t)=(2m)^2(1+\cos\left((2n+1)t\right))$ and $g(t)=(2n+1)^2(1-\cos({2mt}))$. Then \eqref{EqESE} is equivalent to finding a zero of $f-g$ on $(0,\pi)$. Note that both functions are non-negative. 
\begin{center}
    \begin{tabular}{c||c|c}
        $t$ & $0$ &$\frac{\pi}{2n+1}$\\\hline
        $f$ & +&0\\\hline
        $g$ &0&+
    \end{tabular}
\end{center} We know that  $g(\frac{\pi}{2n+1})\neq0$, since zeros of $g$ are integer multiples of $\frac{\pi}{m}$ and $m$ and $2n+1$ are coprime.
Therefore, by the Intermediate Value Theorem $f-g$ has a zero on $(0,\frac{\pi}{2n+1})$ and hence $J$ has ESE.

In addition, if $m=n+1$ then we work with $\cos((2n+1)t)$ and $\cos((2n+2)t)$ whose zeros (and extreme points) are strictly interlacing. Hence,
\begin{center}
    \begin{tabular}{c||c|c|c|c|c|c|c|c|c|c}
        $t$ & $0$ &$\frac{\pi}{2n+1}$&$\frac{\pi}{n+1}$&$\frac{3\pi}{2n+1}$&$\frac{2\pi}{n+1}$&$\frac{5\pi}{2n+1}$&$\ldots$&$\frac{(2n-1)\pi}{2n+1}$&$\frac{\pi n}{n+1}$&$\pi$\\\hline
        $f$ & +&0&+&0&+&0&\ldots&0&+&0\\\hline
        $g$ &0&+&0&+&0&+&\ldots&+&0&0
    \end{tabular}
\end{center}
That is, we get $2n$  open intervals where $f-g$ has exactly one zero which follows from the Intermediate Value Theorem and monotonicity of $f$ and $g$ on these intervals. That is, we can construct a Jacobi matrix with the prescribed even number of ESE occurrences.
\end{proof}

\section{ESE for Persymmetric Jacobi Matrices with Centered Spectra.}

What would work differently if we consider the centered spectrum?
\begin{proof}[Proof of Theorem~\ref{main}]
If $J$ is as in \eqref{J5} and $H$ is a Jacobi matrix realizing PST between end-vertices of the weighted path with centered spectrum $\{c-x,c-y,c,c+y,c+x\}$, then $H=J+cI$ due to Theorem 3 in \cite{H74} and 
\begin{align*}
|\langle e^{-iHt}{\bf e}_0,{\bf e}_0\rangle_{\mathbb{C}^{N+1}}|&=|\langle e^{-i(J+cI)t}{\bf e}_0,{\bf e}_0\rangle_{\mathbb{C}^{N+1}}|
=|\langle e^{-iJt}e^{-ict} I{\bf e}_0,{\bf e}_0\rangle_{\mathbb{C}^{N+1}}|\\
&=|e^{-ict}|\cdot|\langle e^{-iJt} {\bf e}_0,{\bf e}_0\rangle_{\mathbb{C}^{N+1}}|=|\langle e^{-iJt}{\bf e}_0,{\bf e}_0\rangle_{\mathbb{C}^{N+1}}|.    
\end{align*}

That is $J$ and $H$ either both have ESE or both do not. Combining this observation with Claim~\ref{ClaimNoESE} and 
Claim~\ref{ClaimESE} concludes the proof of 
Theorem~\ref{main}.
\end{proof}
%\section{Numerical Examples}

%Pictures with explanations.

%\section{Conclusions}

%\section{Acknowledgments}

\section{Future Directions}
Even though we were able to get a partial result, we observed ESE for higher odd numbers of qubits (see Figure 5).

In \cite{ESE} the authors showed a general approach for all even numbers of qubits and that ESE is impossible for 3 qubits. Note that in the case we considered the occurrence of ESE was contingent on either one of the positive eigenvalues being not an integer multiple of another. Once we scale up the chain to, say, 7 qubits, this notion loses its relevance. Another challenge in handling odd numbers is the presence of the constant term that prevents us from generalizing the approach the authors used in \cite{ESE}.

\begin{figure}[h]\label{moreex}
\caption{Examples of ESE occurrences for chains of 7, 9, and 11 qubits.}
    \begin{center}      \includegraphics[width=0.9\textwidth]{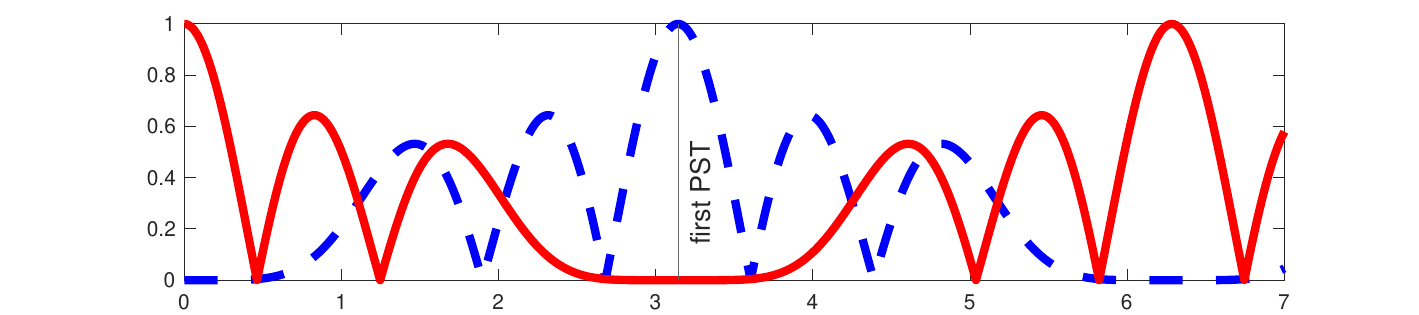}
        
        $7\times7$ symmetric and persymmetric Jacobi matrix with eigenvalues $\{0,\pm3,\pm4,\pm5\}$.\\
                \includegraphics[width=0.9\textwidth]{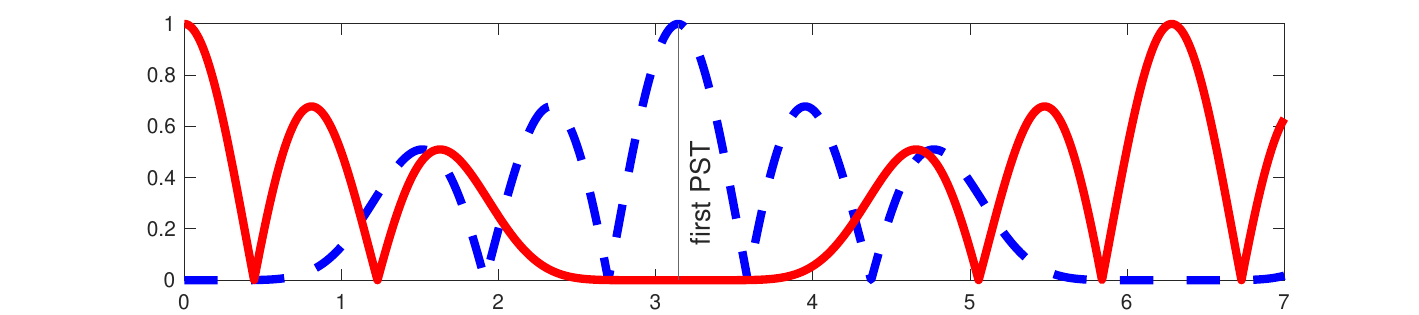}
                
        $9\times9$ symmetric and persymmetric Jacobi matrix with eigenvalues $\{0,\pm3,\pm4,\pm5,\pm8\}$.\\
                \includegraphics[width=0.9\textwidth]{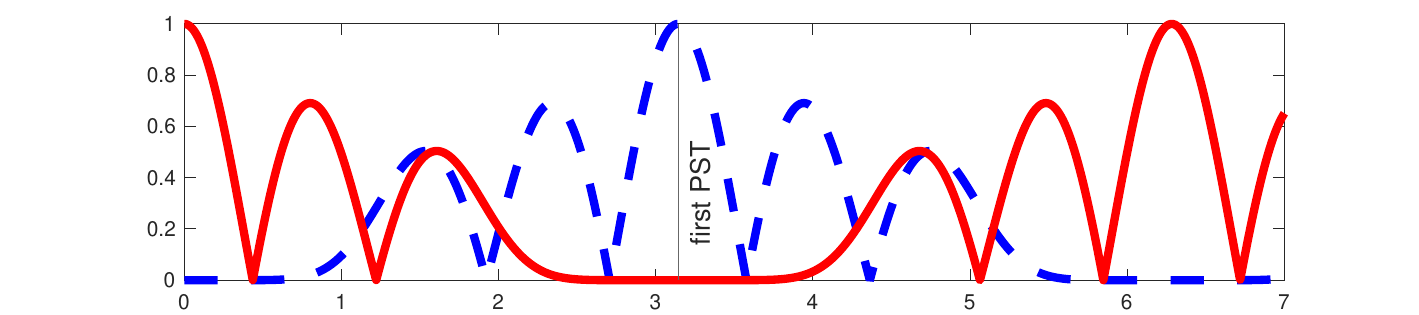}
                
        $11\times11$ symmetric and persymmetric Jacobi matrix with eigenvalues $\{0,\pm3,\pm4,\pm5,\pm8,\pm11\}$.
    \end{center}
    
\end{figure}
%Pictures with explanations.
Thus, the ESE occurrence for all odd numbers greater than 5 remains an open problem.
%\section{Conclusions}

\section{Acknowledgments}
Support for this work was provided by The Women’s Advancement Initiative’s Dorothy Goodwin Scholars Program, which was made possible thanks to a generous bequest from Dorothy Goodwin.  This support does not necessarily imply endorsement by the University of Hartford or The Women’s Advancement Initiative of research conclusions.  A.M. was also partially supported by the Greenberg Junior Faculty Grant and the Coffin Grant from the University of Hartford. Finally, A.M. is extremely grateful to Prof. Christino Tamon and Prof. Maxim Derevyagin for bringing the ESE problem to her attention.

\bibliographystyle{vancouver}
\bibliography{Vancouver.bib}

@Article{	  ESE,
 author	= {Bailey, Rachel and Costa, Sara and Derevyagin, Maxim and Findley, Kai and Zuang, Kai},
title		= {Hamiltonians that realize perfect quantum state transfer and early state exclusion.},
  journal	= {Quantum Inf. Process.},
  volume	= {24},
  number	= {51},
  pages		= {},
  year		= 2025
}

@Article{	  BG,
 author	= {de Boor,C and Golub, G},
title		= {The Numerically Stable Reconstruction of a Jacobi Matrix from Spectral Data.},
  journal	= {Linear Algebra and its Applications},
  volume	= {21},
  number	= {},
  pages		= {245--260},
  year		= 1978
}

@article{B03,
    author = {Bose, S}, 
    title = {Quantum Communication through an Unmodulated Spin Chain},
    journal = {Phys. Rev. Lett.},
    year = 2003,
number=91,
pages=207901
}

@Article{	  DMS,
author={Derevyagin, Maxim and  Minenkova, Anastasiia and Sun, Nathan},
  title		= {A Theorem of Joseph-Alfred Serret and its Relation to Perfect Quantum State Transfer.},
  journal	= {Expo. Math.},
  volume	= {39},
  number	= {},
  pages		= {480--499},
  year		= 2021
}

@article{H74,
    author ={Hochstadt, H.} ,
    title ={On the construction of a Jacobi matrix from spectral data} ,
    journal ={Linear Algebra Appl.} ,
volume ={8},
pages={435--446},
    year = 1974
}

@Article{	  HM,
author={Holley, Alex and  Minenkova, Anastasiia},
  title		= {A Linear Algebra Story: How We Reconstructed a Matrix from its Eigenvalues.},
  journal	= {The College Mathematics Journal },
  volume	= {54},
  number	= {5},
  pages		= {454--463},
  year		= 2023
}

@Article{	 Kay10,
  author	= {Kay, A},
  title		= {A Review of Perfect State Transfer and its
			Application as a Constructive Tool.},
  journal	= {Int. J. Quantum Inf.},
  volume	= {8},
  number	= {},
  pages		= {641--676},
  year		= 2010
}

@article{Kall,
    author = {Kim, S and Monterde, H and  Ahmadi, B and Chan, A and Kirkland,S and Plosker, S} ,
    title ={A generalization of quantum pair state transfer} ,
    journal = {Quantum Information Processing},
 volume	= {23},
  number	= {11},
    year = 2024
}

@Article{	  VZh12,
  author	= {Vinet, Luc and Zhedanov, Al},
  title		= {How to construct spin chains with perfect state transfer.},
  journal	= {Physical Review A},
  number	= {85},
  pages		= {12--323},
  year		= 2012
}

\end{document}